\documentstyle[12pt]{article}
         \makeatletter
         \def\thefigure{\@arabic\c@figure}\def\fps@figure{tbp}
         \def\ftype@figure{1}\def\ext@figure{lof}
         \def\fnum@figure{\protect\footnotesize Fig.\ \thefigure}
         \def\thetable{\@arabic\c@table}
         \def\fps@table{tbp}\def\ftype@table{2}\def\ext@table{lot}
         \def\fnum@table{\protect\footnotesize Table \thetable}
         \makeatother
\begin{document}
\vspace*{0.3in}
\begin{center}
  {\Large \bf
The Photodissociation of $^8B$ and the Solar Neutrino Problem}\\
  \bigskip
  \bigskip
  {\Large  C.A. Bertulani$^{*}$ }  \\
  \bigskip

   Gesellschaft f\"ur Schwerionenforschung, KPII\\
   Planckstr. 1, D-64291 Darmstadt, Germany\\

 \bigskip

  \end{center}

\bigskip
\centerline{\bf ABSTRACT}
\begin{quotation}
\vspace{-0.10in}
The extraction of the photodissociation
cross sections of $^8B$ from
Coulomb dissociation experiments
is investigated.
A careful study is done  on the contributions of the
E1, E2 and M1 multipolarities to the breakup.
A comparison with the
data of a recent experiment is performed.
It is shown that the extraction of the radiative capture
cross sections $^7Be(p,\ \gamma)^8B$ which are relevant
for the solar neutrino problem is not affected appreciably
by Coulomb reacceleration. A non-perturbative model is used
for the purpose.
Emphasis is put on the
perspectives for future experiments which are planned
at the University of Notre Dame, RIKEN (Japan), and
GSI (Germany). An analysis of the total  yields of
``photon-point" processes
in inelastic electron scattering is also done.

\end{quotation}
\vfil
\noindent * Present and permanent address: Instituto de F\'\i sica,
Universidade Federal do Rio de Janeiro\
21945-970 \ Rio de Janeiro, RJ, \ Brazil

\newpage
\baselineskip 4ex

\section{Introduction}

The solar neutrino problem is related
to a factor of two difference between the theoretically
calculated and the experimentally measured neutrino flux from
the Sun. This discrepancy is more accentuated for the measurements
of the Homestake \cite{Da68} and the Kamiokande \cite{Hi89} experiments.
The Gallex \cite{An92} and SAGE \cite{Ab91} experiments also imply
a discrepancy, although more moderate,  between theory and experiment.
While the Gallex and the SAGE experiments measure neutrinos which are
as low in energy as 233 keV, the Homestake and the Kamiokande experiments
measure high energy neutrinos. Thus, the discrepancy seems to be larger
for the higher neutrino energies.

Most of the high neutrinos ($E_\nu >2$ MeV) come from
the beta-decay of $^8B$.
The main mechanism through which the $^8B$ is produced in the Sun is the
$^7Be(p,\ \gamma)^8B$ reaction \cite{Ba89}. If the cross section for
this reaction were half of the value which is presently used as
an input to the Standard Solar Model \cite{Ba89} the Homestake
and the Kamiokande experiments would agree well, although the
gallium results would remain unexplained.
It is also worthwhile to say that
this change might have other implications which can
be in disagreement with the SSM model.
A proposed solution to
the solar neutrino problem is that the discrepancy lies not in the
nuclear cross sections that are involved in the determination of the
neutrino fluxes, but rather in the behaviour of the neutrinos during
their flight to the terrestrial detectors, most likely matter induced
oscillations into other neutrino types (The MSW effect \cite{Ba89}) to
which the detectors are insensitive. However, not all the nuclear cross
sections are known with the desired accuracy, and their inaccuracies will
significantly affect the interpretation of the Davis $^{37}Cl$
(Homestake) and Kamiokande Cherenkov detectors. The reaction
$^7Be(p,\ \gamma)^8B$ accounts for the bulk of the neutrinos seen by
$^{37}Cl$ and is the sole source of neutrinos seen by Kamiokande.

The $^7Be(p, \ \gamma)^8B$ reaction occurs in the
Sun at a temperature corresponding to  a relative
p-$^7Be$ energy of about 20 keV.
Due  to the Coulomb barrier,
the magnitude of the cross section at this energy
is too small to be measured directly. Direct measurements
have been done down to an energy of 120 keV  \cite{Fi83}.
The measurements of this cross section have always been difficult,
because the target is radioactive and relatively short lived ($^7Be$ has
a halflife of 53 days), and the measurements have been plagued with
normalization. Thus, although the most recent and careful
experiments of Fillipone et al. \cite{Fi83} have a quoted error of
10\%, Barker and Spear \cite{BS83} argue that the true result may be
smaller than that used in the predictions of Bahcall \cite{Ba89}, by
as much as 30\%, due to a combination of normalization, data selection
and excitation function uncertainties.

Another controversial point is that
the input in the solar models is an extrapolation
of the cross section to the 20 keV region.
This procedure yields
an astrophysical S-factor close to 20 keV.b \cite{BP92}. These
extrapolations are done based on nuclear models which allow
some flexibility, and only the experiment can tell how
accurate these models are. Therefore, as long as a measurement
of the radiative capture cross section for the reaction
$^7Be(p, \ \gamma)^8B$ at low energies is not done, the solar
neutrino  problem cannot be accepted as solved.

A new technique \cite{BBR86} is now available for
the measurements of these cross sections. One studies the
inverse reaction $^8B+\gamma \longrightarrow ^7Be+p$ through
the Coulomb breakup of $^8B$ on a heavy target. The Coulomb
field of the heavy nucleus provides the equivalent photons.
The $^7Be(p, \ \gamma)^8B$ cross section is obtained by
detailed balance. This method
has been shown to work very well for at least
two reactions \cite{Mo91,Ki91}
(for a recent review see \cite{BR94}).
Experiments
with real photons are infeasible.
But, in first-order, Coulomb
excitation involves the same matrix elements as those of
photo-induced processes. Furthermore,
the coherent field of a large $Z$ nucleus results in very large
dissociation cross sections, which can be studied in present experimental
facilities \cite{BBR86}.
A recent experiment
by Motobayashi et al. \cite{Mo94}
have used the Coulomb dissociation method
to extract the photodissociation cross section
$^8B+\gamma \longrightarrow ^7Be+p$.

The main problems with the method are (a) to
experimentally separate the nuclear from the Coulomb contribution
in the breakup events, and (b) contribution from higher order effects,
usually called by reacceleration, or post-acceleration, effects
\cite{BBK92,BB93,TB93}.
When the nuclear contribution to the breakup cross section  is
large compared to the Coulomb breakup mechanism
for the experimentally measurable kinematical conditions,
the method is useless. This is likely to be the case for the breakup
of oxygen on nuclear targets, in order to extract the much needed
information on the $^{12}C(\alpha, \ \gamma)^{16}O$ relevant
for  the helium burning in massive stars \cite{B94}. For a comprehensive
review on the relevance of the radiative capture reactions in
astrophysics see the paper by Fowler \cite{Fo84}.

The case of Coulomb  breakup of $^8B$ as
a tool to obtain the $^7Be(p, \ \gamma)^8B$ cross section at
low energies has been investigated by many authors recently
\cite{Mo94,TB93,TB94,B94}.
In particular, in ref. \cite{B94}
it was shown that the nuclear contribution
to the breakup of $^8B$ is negligible at forward angles below
a threshold dictated by
the onset of nuclear absorption and diffraction effects.
This is a very useful result, since (in contrast to the
breakup of oxygen) the much simpler semiclassical description of
the Coulomb breakup maybe used.

In this article we consider the Coulomb breakup of $^8B$ at different
bombarding energies, namely 5 MeV/nucl, 50 MeV/nucleon and  250 MeV/nucleon,
e.g., at the Notre Dame University (USA),
RIKEN (Japan) and the GSI (Germany).
Radioactive beams of boron can be obtained at these energies at several
other labs around the world. The advantage of performing a series of
experiments at
different bombarding energies is that the multipolarity content of the
breakup can be disentangled with more accuracy.
This is
shown in section 2.

The reaction $^7Be(p, \ \gamma)^8B$ is dominated
by the electric dipole (E1) multipolarity at very low energies,
below the $1^+$ resonance  at 633 keV. Higher order
effects, or Coulomb reacceleration effects, act so as to distort the
final relative energy of the fragments after the breakup. In ref.
\cite{B94}
this problem was studied by means of a  classical prescription,
based on a Monte-Carlo calculation.
Typel  and Baur \cite{TB94} have recently studied the same problem
within second order time-dependent theory. More recently, Esbensen
et al. \cite{Es94} have shown that while the effect is relevant for
the breakup of $^{11}Li$ projectiles on heavy targets, it is small
for the breakup of $^{11}Be$ and large incident energies.

It has been shown \cite{BB93,Es94}
that a correct treatment of reacceleration
effects can only be accomplished by solving the time-dependent
Schroedinger equation non-perturbatively.  This is because,
the reacceleration is an effect related not only to
the magnitude of the
Coulomb breakup probabilities, which can be small. It is
also an
effect  of distortion (caused by the Coulomb field of the
target) on
the relative motion between the
fragments after the breakup. The absence of such distortions
are of vital importance to validate the Coulomb dissociation
method.
We report in section 3 a
calculation based on the numerical solution of the three-dimensional
Schroedinger equation in a potential model for p+$^7Be$. The model
used is similar to the one used by Esbensen et al. \cite{Es94} for
the breakup of $^{11}Li$ and $^{11}Be$ projectiles. An additional
difficulty here is the inclusion of the Coulomb barrier for the
$p-^7Be$ relative motion.
The results of this non-perturbative model
are compared with previous calculations.
It is shown that the reacceleration effect is much smaller than the
experimental uncertainties that one achieves in the Coulomb dissociation
experiments for the breakup of boron.
The relative population of the $s$, $p$, $d$ and $f$ waves  in
the continuum is obtained. We also deduce the
angular correlations, useful to separate the multipolarity content
of the breakup process.
As a byproduct of our calculations several predictions are made
which are useful for future experiments.

In section 4 we show that
although the luminosity of electron beams are much
higher than the ones obtainable in heavy ion facilities,
it is hard  to use this advantage to
extract the  photodissociation cross sections needed for astrophysics
from inelastic electron scattering.  However, the total yields of
target breakup by inelastic electron scattering are not small for
ideal experiments.
Our conclusions are presented in section 5.

\section{First-order breakup of $^8B$}

The  extraction of the radiative capture cross section
$^7Be(p, \ \gamma)^8B$ from Coulomb breakup experiments
depends on the validity of the first-order perturbation
theory \cite{BBR86}. When this approximation is valid the Coulomb
breakup cross sections of $^8B$ into $p+^7Be$ in
a continuum state with energy
$E$ is given by
\begin{equation}
{d^2\sigma_C\over dE \  d\Omega}
= {8 \over 5}\ \mu_{bc}
\
{1\over (E+Q)^3} \ e^{-2\pi\eta}\ \sum_{\pi\lambda}
{d^2n_{\pi\lambda} \over dE\ d\Omega} (E_x, \ \Omega)
\ S_{\pi\lambda}(E)  \ ,
\end{equation}
where $E_x=E+Q$ is the excitation energy, $Q=137$ MeV  is the binding
energy of the system, and $S_{\pi\lambda}(E)$ is the
astrophysical S-factor
for the electric, $\pi=E$, or
magnetic, $\pi=M$, multipolarity of order $\lambda$.
Detailed balance was used to link the ground state of
$^8B$ ($J^\pi=2^+$) to the proton ($J^\pi={1\over2}^+$)
and the $^7Be$ ($J^\pi={3\over2}^-$) in the final state.
The
functions $n_{\pi\lambda}(E_x, \ \Omega)$ are known as the
virtual, or equivalent,
photon numbers. In ref. \cite{AB89} it was shown that one can calculate
these quantities for all bombarding energies in the semiclassical
approach including the effects of retardation and Coulomb repulsion.
This  is important
since  both retardation and the Coulomb recoil of the classical
trajectory influence the values of $n_{\pi\lambda}
(E_x, \  \Omega)$ at
bombarding energies around 50-200 MeV/nucleon. At lower energies
retardation plays  no role and the results of ref.
\cite{AB89} reproduce the
semiclassical Coulomb excitation theory developed in ref.
\cite{AW56}. At high energies, the Coulomb
recoil is not relevant and the analytical results for the equivalent
photon numbers are also reproduced \cite{AB89}.

The differential cross section $d\sigma_C(E)/dE$ can also be
written in the form of eq. (1), but with the virtual photon numbers
integrated over the scattering angle $\theta$, from zero to
a maximum value  corresponding to a grazing collision.
To display
the multipolarity dependence of the equivalent photon numbers
(angle integrated)
we plot them in figure 1 for the E1, E2 and M1 multipolarities, and
as a function of the excitation energy $E_x$. A lead target is
assumed.
We observe that the
$M1$ multipolarity is about a factor 100, 10 and 3 smaller than the
E1 multipolarity for bombarding energies of 5, 50 and 250 MeV/nucleon,
respectively. This is basically a consequence of the fact that
$n_{M1} \propto (v/c)^2 n_{E1}$, where $v$ is the projectile velocity.
The E2 multipolarity is about 2-3 orders of magnitude larger than the
E1. This is  a reflection of the  ``tidal" component of an $1/r^2$
force, which is more  noticeable for small Fourier components of
the field, and small bombarding energies.

The astrophysical S-factors for the $^7Be(p, \ \gamma)^8B$
reaction is dominated by an $1^+$ resonance at E=633 keV. At  lower
energies the  reaction is dominated by the E1 capture from s and
d waves in the continuum.
A model calculation by Kim et al. \cite{Ki87} yields an
E2 cross section which
is a factor 500-1000 smaller than the E1 cross
section,
over the
relevant energy domain. In view of the results presented in figure
1 we conclude that the  E2 breakup maybe as important as
the E1 breakup  mode in collisions with the bombarding energies
studied here. This fact has been exploited in ref. \cite{LS94}
to suggest that the data reported by Motobayashi et al. \cite{Mo94}
would imply  an $S_{E1}$-factor which is about  30\% smaller than what
was extracted from the experimental analysis.
In view of the potentiality  of the Coulomb dissociation method
for this particular problem it is important to look at
this question more closely.

To calculate $d\sigma/dE$ we will use the results of Kim et al. \cite{Ki87}
for the astrophysical S-factors of the  E1, E2 and  M1
capture modes. The M1 and E2 capture modes are influenced by
the continuum resonance at 633 keV. This influence is
much stronger in the M1 case, so that this capture mode is
only relevant for the energy region around 633 keV.
In figure 2  we plot the energy spectrum of the Coulomb
breakup of $^8B$ projectiles on lead at several energies and
for the E1, E2 and M1 dissociation modes. Apart from a small
resonant component in the E2 dissociation mode, the E1 and E2
contributions to the breakup are dominated by a prominent
bump  around 300-500 keV in the relative energy of the
fragments. This peak is the result of the folding of a
increasing function of $E$, the Sommerfeld penetration factor
$\exp\{-2\pi\eta\}$, with a decreasing
function of $E$, the equivalent photon numbers times the
$(E+Q)^{-3}$ factor in eq. (1).
It is also clearly  seen that the E2 component of the Coulomb
breakup is comparable to  the E1 component at the region of
astrophysical interest ($E<1$ MeV).  However, this result is
strongly dependent on the angular kinematical conditions in
measurements of the breakup cross sections, and also on the
model for the capture cross sections used as input.

To show  this more clearly we plot in figure 3 the  angular
distribution of the c.m. of the fragments with
a relative energy of 100 keV, and for the E1 and E2 multipolarities.
A peculiar feature  of the  angular distribution is that
the E1 breakup mode is concentrated at small scattering angles,
while the E2 breakup mode is spread over a large angular region.
This is more apparent in figure 4, for $E=500$ keV.

\bigskip
\bigskip
\begin{center}
\begin{tabular}{|l|l|l|l|l|r|} \hline\hline
$E_{lab}$  [MeV/nucl.] &$\sigma (\theta<\theta_m)$
[mb]&&$\sigma(\theta<\theta_S)$ [mb]& \\ \hline
&E1&E2&E1&E2 \\ \hline\hline
5&572&392&159&73 \\ \hline
50&437&210&172&23 \\ \hline
250&178&66&38&0.95 \\ \hline\hline
\end{tabular}
\end{center}
{\bf Table I} -  E1 and E2 components in the Coulomb breakup
of $^8B$ for events below a maximum scattering angle of
$\theta_m=30^\circ$, $3^\circ$ and $0.3^\circ$ or
$\theta_S=75^\circ$, $9.4^\circ$ and $2.5^\circ$
and at bombarding energies corresponding to 5, 50 and
250 MeV/nucleon, respectively.
\bigskip
\bigskip

This important feature allows one to select events within an
angular region and to manipulate the relative contribution
of the E2  and E1 components to the breakup. For example,
by selecting events for which the  scattering occurs
below a certain maximum angular value the E2 contribution can be
reduced  considerably.
The calculations of the  cross sections presented in figures 2-4
were obtained by integrating the equivalent photon numbers in
eq. (1) from $\theta=0^\circ$ up to $\theta=\theta_S$, where
$\theta_S$ is  the  grazing scattering angle, above which the
strong interaction sets in. These are
$75^\circ$, $9.4^\circ$, and $2.5^\circ$ for $E_{lab}=5$,
50 and 250 MeV/nucleon, respectively. Let us now assume that
in a given experiment one selects events which occur below the
angles of $30^\circ$, $3^\circ$ and $0.3^\circ$
for $E_{lab}=5$,
50 and 250 MeV/nucleon, respectively. Then the total cross
section, integrated over angle and energy, varies considerably
and is shown in table I. We notice that the E2 contribution is
much more affected by the angular limitation  of the
events than the E1 component.

We now compare our results with the experiment of Motobayashi et
al. \cite{Mo94}
for the angular distribution of the breakup of
$^8B$ on lead at 46.5 MeV/nucleon. We use the
detection efficiency for the data with $E_{rel}=600$ keV
obtained from the authors \cite{Ga94}.
In the experiment the
uncertainty  of the  relative energy
of the fragments amounted to 100 keV. Thus, we used as
input in the calculations the S-factors averaged over
energy bins of 100 keV and the model of Kim et al. \cite{Ki87}.
However, for the E1 capture mode we multiplied the results of
Kim et al. \cite{Ki87} by a factor so that we could reproduce
the data from ref. \cite{Mo94}. After this procedure, the
effective $S_{E1}$-factor that we use in our calculations
is close to 18 eV b.
Our results are compared with the data of Motobayashi  et  al.
\cite{Mo94} in the figure 5. The solid line is the breakup
cross section
multiplied by the detection efficiency assuming only the E1
breakup mode. The dashed line includes the E2 breakup mode.
In the experiment of Motobayashi et al. the restriction
on the angular distribution of the fragments considerably
decreased the effects of the E2 contribution to  the breakup.
But,
if the model of Kim et al. \cite{Ki87} is valid they might not
be negligible, as suggested in ref. \cite{LS94} and
concluded from figure 5.
However, one must be skeptical about any conclusions
related to the E2 contribution in this reaction. The
model of  Kim et al. \cite{Ki87} has been criticised
by Barker \cite{Ba80}. In several other models
\cite{Ba80,DB94,TB94} the
E2 and M1 contribution to the $^7Be(p, \ \gamma)^8B$
reaction is considerably smaller than the ones
predicted by the model of Kim et al. \cite{Ki87}.
If we use the model
of Typel  and  Baur \cite{TB94}
for the E2 capture model of
the $^7Be(p, \ \gamma)^8B$
reaction we get the dotted line in figure 5, which
shows almost no influence of  this multipolarity to the
data of ref. \cite{Mo94}. A recent discussion on the
contribution of the E2-breakup to the data presented in
fig. 5 has also been published in refs. \cite{LS94,GB95}.

We notice that the pattern of the theoretical angular distribution
shown in fig. 5 is slightly different than the one shown in fig. 2 of
ref. \cite{Mo94}, particularly at small angles.
The reason for these differences
is not quite understood, since we use the efficiency curve suplied by
those authors. A possibility is that in ref. \cite{Mo94}
the efficiency curves
were obtained by using a theoretical prediction for the dissociation
modes as input to a Monte-Carlo simulation in which the detection
efficiency and the geometry of the detectors were accounted for.
This contrasts with the approach used here in which a simple folding
of the theoretical prediction and the ``model-dependent" efficiency
curve is done.
It has also been claimed
\cite{Mo95} that the
angular averaging is not so simple as treated here, because of the
poor experimental resolution and rather large energy bin sizes.
It seems that this averaging,
with proper inclusion of the detector
geometry, has been made more accurately in ref.
\cite{Mo94} with the Monte-Carlo simulation.

By switching
off  the retardation effects in the calculation of the virtual
photon numbers\footnote{This can be done by
neglecting retardation effects in the calculation of  the
electromagnetic propagator, and on the calculation of  the
Coulomb trajectory \cite{AB89}.} which enter eq. (1)
we obtain an average of 3 percent decrease
of the calculated curves presented in figure 5, a
small effect for this bombarding energy.
But, these
effects will increase with the bombarding energies and
for higher excitation energies, and should be considered
in a more detailed analysis of future experiments.
For the other (higher) relative energies presented in ref.
\cite{Mo94} we find that
the ratio $\sigma(E1+E2)/\sigma(E1)$
reduces considerably. In view of the
discrepancies between the different nuclear models
for the $^7Be(p, \ \gamma)^8B$
reaction we feel that no quantitative conclusions
can be drawn about the amount of the M1 and the
E2 contribution to the Coulomb breakup of $^8B$. In fact,
the Coulomb dissociation experiments will certainly  be
very useful to solve this problem, since the
several multipolarities contribute differently at
different kinematical conditions.

In figure 6 it is shown the energy spectrum of the
breakup yields and compared with the experimental data
of Motobayashi et al. \cite{Mo94}.
For qualitative comparisons we use the model of Kim et al.
\cite{Ki87}. Again, we scale down the E1 capture cross section
from Kim et al. \cite{Ki87} which amounts to use a constant
factor $S_{E1}=18$ eV b.
The solid curve in fig. 6 is our theoretical
calculation (E1 only) folded with the detection efficiency.
The dashed curve includes the E2 contribution to the breakup.
In this case the same detection efficiency is used
to obtain the E2 component. However, the E2 cross
section was integrated only over the angular region where the
detection efficiency (figure 2 of ref. \cite{Mo94}) is
relevant. As stated before (see figures 4 and 5), the exclusion of
large angles, reduces considerably the E2 contribution to
the  breakup data presented in ref. \cite{Mo94}.
Thus, it is not correct to assume the same detection efficiency for
the E2 and the E1  breakup mode. If we assume that the
E1 and E2 efficiencies are the same, fig. 6 shows that
only the point at 600
keV is appreciably modified by the inclusion of the  E2 and of the
M1 components.
This  is basically due to the presence of
the $1^+$ resonance at 633 keV.

We notice that the $S_{E1}=$18 eV b value used in our calculations
contrasts with that used ($S_{E1}=15$ eV b) in the calculation of ref.
\cite{Mo94} and presented in their fig. 2. The reason for this
difference might be due to their different numerical approach,
as we discussed above in connection with the angular distribution
presented in fig. 5. However, we note that the $S_{E1}=18$ eV b
adopted here seems consistent with the (very preliminary) value
of $S_{E1}= 16.7 \pm 3.2$ eV b given by those authors, and claimed to
be the best fit to their full set of data.

In figure 7 we show the $\sigma_{E2}/\sigma_{E1}$ ratio as
a function of the  scattering angle, the laboratory
energy, and for relative energies between the fragments
equal to 100 keV (a) and 500 keV (b). Once again, the model of
Kim et al. \cite{Ki87} was used.
One sees that the
ratio decreases with the relative energy and with scattering
angle.
The ratio is also
smaller for bombarding energies in the range of 10-200 MeV/nucleon.
Experiments at different laboratory energies would allow a
separation of the relative contributions of the E1 and E2
multipolarities to the breakup.

As stated before, the validity of the Coulomb dissociation
method to extract the astrophysical S-factors is strongly
dependent on the validity of the first-order perturbation
theory. In the next section we present
a study of a non-perturbative approach to reacceleration
effects.

\bigskip\bigskip

\section{Non-perturbative Breakup of $^8B$}

A non-perturbative approach to the study of reacceleration
effects was proposed in refs. \cite{BB93,Es94}.
In the system of reference of $^8B$ we can write the
time-dependent wavefunction for the $p+^7Be$ relative
motion as
\begin{equation}
\Psi({\bf r}, \ t)={1\over r} \ \sum_{lm}
\ u_{lm} ({\bf r}, \  t) \ Y_{lm}(\hat{\bf r})
\end{equation}
Initially the $^8B$ is assumed to  be in its bound state.
We use a static $p+^7Be$ Woods-Saxon
nuclear potential which reproduces the binding energy of $^8B$.
The parameters used are $V_0=-32.65$ MeV, $R= 2.95$ fm and
$a= 0.52$ fm for
the depth, mean radius and diffusivity of  the
potential, respectively. The Coulomb potential of a uniform
distribution of the $^7Be$ charge with radius $R_C=R$ is
added.

We will treat the dynamical
problem of the evolution of the wavefunction of the
system with a time-dependent Coulomb interaction between
the projectile and the target.
However, we keep the nuclear potential constant.
It is well known that a single set of potential parameters
cannot reproduce simultaneously the binding, as well as
continuum resonances of a system (see, e.g., the work of
Tombrello
\cite{To65}).
But, these are small changes in the potential parameters, and
since we are interested here in the reacceleration caused
by the time-dependent Coulomb interaction, we will neglect the
dynamical  modification of the nuclear potential.

For simplicity we will assume  that the projectile moves
along a straight-line. To account for retardation effects we
use the Lienard-Wiechert interaction between the target and
the projectile \cite{Ja75}.
Expanding the interaction in lowest order
at the region of the projectile
we find that the time-dependent
E1 potential is given by
\begin{equation}
V_{E1}(t)=\sqrt{2\pi \over 3} \  Ze\gamma \ r \ \biggl\{
\Big[ Y_{1,-1}({\bf r})-Y_{11}({\bf r})\Big] \ b +
\sqrt{2}\ v \ t \ Y_{10}({\bf r})\
\bigg\} \ f(t) \ ,
\end{equation}
where
\begin{equation}
f(t)=\Big[b^2+\gamma^2v^2t^2\Big]^{-3/2}
\ .
\end{equation}
In the equations above $\gamma=(1-v^2/c^2)^{-1/2}$, $Z$ is the
target charge, and $b$ is the impact parameter.
Eq. (3)  is the effective E1
interaction. It is obtained by using the form
($\rho \phi - {\bf j.A}/c^2$) for the electromagnetic interaction
and the continuity equation to link the magnetic and the
electric interactions. This is important for collisions where
$v/c$ is not negligible. The effects of retardation is manifest in
the appearance of the factors $\gamma$ in eq. (3).
We will restrict ourselves here to the  E1
interaction. But our conclusions will  be of general validity
and will not be affected by the use of this
approximation.

When the expansion (2) and the nuclear plus the
time-dependent interaction $V_{E1}$ is inserted in the
time-dependent Schr\"odinger equation
a set of coupled differential equations is obtained. The
expansion is truncated at $l=4$, for practical purposes,
and the coupled equations is solved numerically.
The numerical procedure is similar to that explained in ref.
\cite{BB93}.
For example, if we restrict only to the time-dependent E1
interaction, the equation to be solved is
\begin{eqnarray}
&&\bigg[ {d^2 \over dr^2} -{l(l+1)\over r^2} -
{2 \mu_{bc}\over \hbar} \ V_N(r) \bigg] \ u_{lm} (r, \ t)
- {2 \mu_{bc}\over \hbar} \ {(-1)^m\over \sqrt{2}}\
f(t) \\ \nonumber
&\times & \sum_{l'm'} \sqrt{(2l+1)(2l'+1)}
\ \biggl({l \atop 0}{1\atop 0}{l '\atop 0} \biggr)
\Biggl\{ i b \biggl[
\biggl( {l \atop -m}{1\atop 1}{l '\atop m'} \biggr)
-\biggl( {l \atop -m}{1\atop -1}{l '\atop m'} \biggr)
 \biggr]\\ \nonumber
&+& \sqrt{2} \ v \ t \
\biggl({l \atop -m}{1\atop 0}{l '\atop m'} \biggr)
\Biggr\} u_{l'm'} (r, \ t)
=-{2\mu_{bc} \over \hbar} \ {\partial u_{lm} \over \partial
t}\ ,
\end{eqnarray}
This equation couples $u_{lm}$ with neighbouring angular
states (due to the discrete values of the  Wigner symbols)
with $\Delta l=\pm 1$ and $\Delta m=\pm 1,0$.
Including the E2 interaction, complicates a bit more
the coupled equations, but they are  treatable with a
truncation in the maximum $l$-value.

We use a mesh of 1000 points for the coordinate $r$, with a
mesh size of
$\Delta r=0.1$ fm. The time is discretized in a mesh of 800
points with a mesh size of 1 fm/c, starting at $t=-200$ fm/c.
As explained in refs. \cite{BB93,Es94},
this numerical approach allows one
to calculate the time dependence of the wavefunctions and
observables useful to study the effects of reacceleration,
e.g., occupation probabilities, momentum and energy shifts, etc.
We will  concentrate here on the results of this
numerical approach which are relevant to  the $^8B$
breakup. We will first assume a bombarding energy of
50 MeV/nucleon.

The time-dependent wavefunctions computed at 600 fm/c are
not influenced by the Coulomb field of the target and are  running
waves (time-dependent continuum states) added to the remaining
part of the
ground state wavefunction (which is an $l=1$  state).
The continuum states are obtained by removing
the ground-state part of the time-dependent wavefunction , i.e.,
\begin{equation}
\Psi_c=\Big[ \Psi(t)-<\Psi(t)\Big| \Psi_0>
\ \Psi_0 \Big]\ \Big[ 1- |<\Psi_0|\Psi(t)>|^2\Big]^{-1/2}
\ .
\end{equation}

In figure 8 the probability density of the  $p+^7Be$
relative motion in the continuum states is plotted as
a  function of the radial distance, at t=400 fm/c,
and for a collision with a lead target with an
impact parameter $b=15$ fm.
Also shown is the ground state wavefunction. One
sees that at this time the continuum wavefunction
is already far from the range of the static nuclear potential.
It contains the characteristic behaviour
of a spreading wavepacket.

The wavefunction presented in figure 8 is an
admixture of several continuum angular momentum states.
In figure 9 we show the occupation probabilities
of different continuum angular momentum states, for
a collision with an impact parameter of $b=15$ fm and
$b=50$ fm, respectively. We observe that at $b=15$ fm
the states
are more evenly occupied than at b=50 fm. This occurs
because at b=15 fm the interaction is stronger and
the continuum-continuum coupling is also stronger. The most
relevant states are $l=0,1,2,3$. At $b=50$ fm the
interaction is weaker, and
the continuum-continuum coupling is small.
In this case, first order perturbation theory should work
well. In fact, at large impact parameters the
most occupied states are the s-wave ($l=0$) and the
d-wave ($l=2$) states.
These are the same states which are relevant
for the dipole capture from the continuum of the  $p+^7Be$
system \cite{Ki87}.

As stated above,
of interest here is the comparison between the numerical results
with the first-order perturbation theory. Only then we can access
the magnitude of the reacceleration effect. In  first-order
perturbation theory the excitation amplitude
is given by
\begin{equation}
a_{fi}(\omega) = {1\over i\hbar} \ \int_{-\infty}^\infty
e^{i\omega t}\ <f\Big|V_{E1}(t)\Big|i>
 \ ,
\end{equation}
where $\hbar \omega=(E_f-E_i)$ is the excitation energy.
The time integrals can be done analytically.
The amplitude is proportional to the
electric dipole and quadrupole matrix-elements for a
transition between the ground state and a
continuum state with energy $E=\hbar \omega -Q$.
The continuum states are determined numerically for the same nuclear
potential. We select the $l= 0-3$
continuum states as input
to calculate the matrix elements.

To compare with the non-perturbative calculations, we
Fourier transform the time-dependent wavefunction. One
gets
\begin{equation}
\Psi_c({\bf p})=\sum_{lm} {\cal C}_{lm} (p) \ Y_{lm}
(\hat {\bf p}) \,
\end{equation}
where
\begin{equation}
{\cal C}_{lm}(p)=\sqrt{2 \over \pi} \
i^l \ \int dr \ r \ j_l(pr) \  u_{lm} (r, \ t)
\end{equation}
where $j_l$ is the spherical Bessel function and
$u_{lm}(r, \ t)$ are the radial
continuum wavefunctions, with
the ground-state subtracted.
The probability density for an excitation
to a final state with energy $E$ is directly
obtained from the above result. It is
\begin{equation}
{\cal P}(b, \ E, \ \Omega )= {1\over 2}
\ \Big( {2m_{bc} \over \hbar^2} \Big)^{3/2} \
\sqrt{E} \ \Big|\Psi_c({\bf p})\Big|^2 \ .
\end{equation}
Integrating over $\Omega$ we get
\begin{equation}
{\cal P}(b, \ E )=
{1\over 2}
\ \Big( {2m_{bc} \over \hbar^2 }\Big)^{3/2} \
\sqrt{E} \ \ \sum_{lm} \Big|{\cal C}_{lm}(p)\Big|^2
\ .
\end{equation}

The form of the energy spectrum of the relative motion
of the fragments in first-order perturbation theory (solid)
and in the  non-perturbative approach (dashed) is shown in
figure 10.
At small impact parameters
the coupling between the continuum states is stronger
and the reacceleration effect changes the form
of the spectrum. Due to the reacceleration effect higher
energy states are populated than in the perturbative
calculation.
At large impact parameters this effect
is smaller, as clearly seen in fig. 10(b).

The average energy of the  relative motion can be compared
with the results of the first-order perturbation theory. This can be
obtained by using eq. (8)  to obtain the first-order
energy spectrum and the non-perturbative one described above.
Averaging over impact parameter, the result is  shown in
figure 11 for the $^8B+^{208}Pb$ reaction as a function of
the bombarding energy (dashed line).

In ref. \cite{BBK92} a simple semiclassical  model was used to
describe the reacceleration effect. Neglecting the binding
energy of the ($a=b+c$) system, the reacceleration causes a
shift in the energy of the fragments which is given
in the model of ref. \cite{BBK92} by
\begin{equation}
\Delta E_b=-\Delta E_c={\pi \over 4} \
\Big( {Z_b \over Z_a}-{m_b  \over m_a}\Big) \
{Z_aZe^2\over b}
\ .
\end{equation}
These energies are much smaller than the beam
energy $E_{lab}$. The energy shifts can thus be
calculated as small corrections to the velocities
of the fragments. The modification of the
final relative motion energy of the fragments due to
the reacceleration effect is, in this approximation,
equal to
\begin{equation}
\Delta E_{rel}={1\over 2}\ \mu_{bc}\ (v_b-v_c)^2 \simeq
{1\over 4} \ {m_a \over \mu_{bc}} \
{(\Delta E_b)^2 \over E_{lab}} \ .
\end{equation}
The results obtained by using this approximation is also
shown in figure 11 (solid-line) after an average over
the  impact parameter  (this  average is  weighted by the
breakup  probabilities, as explained in ref. \cite{BBK92}).
We observe that, despite the very different approaches
used, the agreement between the two calculations is
quite reasonable. Also shown in figure 11 is the Monte-Carlo
method used in ref. \cite{B94}  to  calculate the
reacceleration energy from a classical prescription
(solid circles). The
error bars are inherent to numerical accuracy. We see
that the numerical approach presented here yields the smallest
result for the reacceleration energy (dashed-line). This can
be understood as follows. In the approaches of refs.
\cite{BBK92,B94} the reacceleration energies were calculated
by assuming that the breakup occurs at the distance of
closest approach. This overestimates  the reacceleration
energy, since  in a quantum mechanical approach the proton
is first brought to the continuum, then tunnels the Coulomb
barrier, inducing  a time-delay to the reacceleration
process. This leads to
an effective breakup position which is farther away than
the distance of closest approach. This is more
critical  in the approach used in ref. \cite{B94}.
In any case, the reacceleration effect poses no major
restriction to the Coulomb dissociation of $^8B$.

In ref. \cite{Es94} it was shown that the reacceleration effects
are very important for the breakup of $^{11}Li$ projectiles on lead
at an incident energy of 28 MeV/nucleon.
However, it was also shown that the effect is negligible for
the breakup of $^{11}Be$ on lead at 72 MeV/nucleon. This result was
ascribed to the higher binding of $^{11}Be$ and to the
smaller effective dipole
charge \cite{Es94}. For the breakup of $^8B$ projectiles, the presence
of the Coulomb barrier diminishes further the effect of reacceleration.
Our results were obtained using a potential model for $p+^7Be$,
similar to that of Kim et al. \cite{Ki87}. As, discussed earlier, this model
is thought to be very poor and unreliable \cite{Ba80}.
The use of a more refined model for the purpose of calculation of
the effects of reacceleration is very difficult and beyond the scope of this
work.

Finally, we calculate the angular correlation of the
fragments in the frame of reference of the
$p+^7Be$. This  could be a useful tool to disentangle
the  E1 from the E2 dissociation mode in future experiments.
In our model
we can study this effect by varying the impact parameter,
since as we have seen in figure 9, the occupation
probabilities for L=1 and 3 states (accessible also via
direct E2 transitions) are also populated via the E1
interaction at small impact parameters.
The angular correlation is just proportional to
$|\Psi({\bf p})|^2$, obtained from  eqs. (8) and (9).
In figure 12 we show the angular correlation of the
fragments, in the frame of reference of the $^8B$
projectile, for $E=500$ keV,
$\theta=4^\circ$, a
bombarding energy of 50 MeV/nucleon, and
$b=15$  fm (dashed) and $b=50$  fm (solid). Although
small, the  occupation of the l=1, and 3 states cause
sizeable modification of the angular correlation. The
simpler approach of Baur and Weber \cite{BW89} can be
used due to  the validity of the first-order perturbation
theory.

\bigskip

\section{\bf Photodissociation by Electron Beams}

The luminosity of electron beams are
so much higher than the ones at heavy ion beam
facilities that
electron scattering experiments should in principle
be another possible
tool of obtaining the photodissociation
cross sections useful for astrophysical purposes.
While such procedure would be useless
for the $^8B$ case, it could be useful for other radiative
capture reactions of interest, e.g., the
$^{12}C(\alpha, \ \gamma )^{16}O$ reaction.

It is well known that for small momentum transfers electron
scattering probes the same matrix elements as in
photonuclear experiments (see, e.g, \cite{EG}).
The condition is that the
four-momentum
transfer is much smaller than electron four-momentum
(we use now $\hbar=c=1$), i.e.,
$q_\mu=(k'-k)_{\mu} \ll k_{\mu}$. Also, for forward scattering,
the longitudinal  momentum transfer is
\begin{equation}
(k'-k)_3=k\ \cos \theta - k \simeq
k'-k \simeq {E'-E\over  v} \simeq
{\omega \over c} \ ,
\end{equation}
where $k_3$ is the component of the electron momentum
along the beam axis.
Thus, for the scattering at very forward angles, $q\equiv
|{\bf k'-k}|
\simeq \omega /c$, i.e., the momentum transfer is the same
as the one by a real photon. Under these circumstances, and
in the long-wavelength approximation, both the transverse
and the Coulomb matrix elements for inelastic electron scattering
are proportional to the electromagnetic matrix elements \cite{EG}.
It is easy to show that
the differential cross section for electron scattering
becomes in this limit
\begin{equation}
{d^2 \sigma_e \over d\Omega \ d\omega}
={1\over \omega} \ \sum_{\pi\lambda}
{dn_{\pi\lambda} \over d\Omega} \
\sigma^{\pi\lambda}_\gamma (\omega)
\end{equation}
where $\sigma_\gamma^{\pi\lambda}$ is  the photonuclear
cross section, and
\begin{eqnarray}
{d n_{M\lambda} \over d\Omega}&=&
{\alpha \over \pi^2} \
\Big[ {EE'-(\hat{\bf Q}.{\bf k})(\hat{\bf Q}.{\bf k'})
-m^2 \over q^4} \Big] \\ \nonumber
{d n_{E\lambda} \over d\Omega}&=&
{\alpha \over \pi^2} \
\bigg\{ \Big( {\lambda \over \lambda+1}
\Big)
\Big[ {EE'+{\bf k.k'}+
m^2 \over Q^4} \Big] +
\Big[ {EE'-(\hat{\bf Q}.{\bf k})(\hat{\bf Q}.{\bf k'})
-m^2 \over q^4} \Big] \bigg\}
\end{eqnarray}
where $Q^2=2EE'\sin^2 (\theta/2)$, $q^2=Q^2-\omega^2$, and
$\omega=E-E'$.

In order to guess the reaction yields due to the electron
scattering at the ``photon-point" ($q=\omega$), we integrate the
above equations
over angles. Since they are a rapidly decreasing
function of the scattering angle, we extend the integration
to all angles to obtain an upper limit of the photon-point
physics in electron accelerators.
In this approximation the differential
cross section $d\sigma_e/d\omega$ can  be expressed
in the same form as eq. (15), but with the total
virtual photon numbers given by
\begin{eqnarray}
n_{M\lambda} (\omega)&=&
{2 \alpha \over \pi} \
\Big[ \ {\rm ln}\xi -{1\over 2} \Big] \\ \nonumber
n_{E\lambda}(\omega) &=&
{2 \alpha \over \pi} \
\Big[ \ {\rm ln} \xi -{1\over 2}+{\lambda \over \lambda+1}
\Big] \  ,
\end{eqnarray}
where $\xi=2\gamma E/\omega$.

There are two important differences between this result  and the
virtual photon numbers that one obtains from Coulomb
excitation experiments. Firstly, the virtual photon numbers carry
a factor $Z^2$ due to the coherence effect of
all charges in the nucleus. Secondly,
the Compton wavelength of a nucleus is much smaller than that
of an electron. The ``smearing" of the electron wavefunction in
the direction perpendicular to the beam weakens the strength
of its interaction with the nucleus (see a discussion of this
effect in ref. \cite{Ja75}).
Thus,
the cross sections for photo-dissociation in electron
scattering experiments are much smaller than in Coulomb
excitation. This disadvantage could be compensated by
means of a large beam luminosity.

The cross section for the reaction
$^{16}O(\gamma, \ \alpha )^{12}C$  is about 3 nb for
a relative energy of 1 MeV. Using the eqs. (18) we
get $\sigma_e^{E1}\simeq\sigma_e^{E2} \simeq 1$ nb
for CEBAF energies ($E \sim 4$ GeV). For a beam
luminosity of $3 \times 10^{38}$ $cm^2/s$ a considerable
number of events can be obtained with a fixed target
experiment. However, the extraction of very slow
reaction products from the target poses a serious
experimental difficulty. Targets of $^8B$ are impossible
to construct due to the small lifetime ($\sim 700$ ms).
However, an attempt to use this method to study the
$^{12}C(\alpha, \ \gamma)^{16}O$ would be worthwhile.
As shown in ref. \cite{Di87}, coincidence experiments with the
identification of the final products, i.e., the $\alpha$ and/or the
oxygen, are quite possible. For astrophysical purposes a
triple coincidence measurement is necessary.

\bigskip

\section{Conclusions}

The Coulomb dissociation of $^8B$ is a promising tool
to extract the radiative capture cross section
$^7Be(p, \ \gamma)^8B$ at low energies. The E1 and
E2 contribution to the breakup are separable by using
proper kinematical conditions. Instead of posing a
difficulty, the enhancement of the E2 contribution to the Coulomb
breakup at low energies is another useful tool to
access information on the relative importance of the many
multipolarities which might contribute to a certain
radiative capture reaction, and which otherwise can only be
extracted from dubious theoretical models.

The effects of reacceleration
are small and can be neglected.
We have shown this with a series of different approaches,
comparing perturbative and non-perturbative methods.

\bigskip\bigskip

I indebted to
Drs. H. Esbensen, J. Bahcall and K. Langanke for useful discussions.

\newpage
\noindent
{\bf Figure Captions}  \\

Fig. 1 \,\,\,
The multipolarity dependence of the equivalent photon numbers
(angle integrated)
for the E1 (solid), E2 (dashed) and M1 (dotted) multipolarities, and
as a function of the excitation energy $E_x$. A lead target was
assumed. Figure (a), (b) and (c) is for $^8B$ projectiles with
bombarding energies of 5, 50 and 250 MeV/nucleon, respectively.
\bigskip

Fig. 2 \,\,\,
The energy spectrum of the Coulomb
breakup of $^8B$ projectiles on lead at
5 (a), 50 (b) and 250 MeV/nucleon (c) and
for the E1, E2 and M1 dissociation modes.

\bigskip

Fig. 3 \,\,\,
Angular
distribution of the c.m. of $p+^7Be$ with
a relative energy of 100 keV, and for the E1 and E2 multipolarities.
Figure (a), (b) and (c) are for bombarding energies of
5, 50 and 250 MeV/nucleon, respectively.

\bigskip

Fig. 4 \,\,\,
Same as in fig. 3, but for $E=500$ keV.
\bigskip

Fig. 5 \,\,\,
Angular distribution of the breakup of
$^8B$ on lead at 46.5 MeV/nucleon.
The data points are from Motobayashi et al. \cite{Mo94}.
The solid line is a theoretical calculation for the E1
component folded with the detection efficiency. The dashed
and dotted curves include the E2 component with different
inputs of the E2 capture in the $^7Be(p, \ \gamma)^8B$
reaction.

\bigskip

Fig. 6 \,\,\,  The energy spectrum
of the breakup of
$^8B$ on lead at 46.5 MeV/nucleon.
The data points are from Motobayashi et al. \cite{Mo94}.
The solid line is a theoretical calculation for the E1
component folded with the detection efficiency. The dashed
curve include the E2 component using the model of Kim et
al. \cite{Ki87} as input for
capture cross section $^7Be(p, \ \gamma)^8B$.

\bigskip

Fig. 7 \,\,\, The ratio $\sigma_{E2}/\sigma_{E1}$ as
a function of the  scattering angle and of the laboratory
energy, and for relative energies between the fragments
equal to 100 keV (a) and 500 keV (b).

\bigskip

Fig. 8 \,\,\, The probability density of the  $p+^7Be$
relative motion in the continuum states as
a  function of the radial distance and  at t=400 fm/c.
The reaction $^8B+Pb$ at 50 MeV/nucleon is  considered.

\bigskip

Fig. 9 \,\,\, The occupation probabilities
of s (l=0), p (l=1), d(l=2),
f  (l=3) and g (l=4) waves in the  continuum, for
a $^8B+^{208}Pb$ collision at 50 MeV/nucleon and
an impact parameter of $b=15$ fm (a) and
$b=50$ fm (b).

\bigskip

Fig. 10 \,\,\, Energy spectrum of the relative motion
of $p+^7Be$ in first-order perturbation theory (solid
line)
and in the  non-perturbative approach (dashed) at
(a) b=15  fm and (b) b=50 fm. A $^8B+^{208}Pb$
collision at 50 MeV/nucleon is  assumed.

\bigskip

Fig.  11 \,\,\, Total energy associated with the
reacceleration effect in the Coulomb breakup $^8B$
on lead targets, as a function of the laboratory
energy per nucleon.

\bigskip

Fig. 12 \,\,\,  Angular correlation between the proton and
the $^7Be$-nucleus in their c.m. frame of reference in the
Coulomb breakup of $^8B$ on lead at 50 MeV/nucleon
and a scattering angle of $4^\circ$.

\end{document}